\documentclass[12pt]{article}

\usepackage{array}
\usepackage{epsfig}
\usepackage{amssymb}
\usepackage{graphics,graphpap}

\renewcommand{\thefootnote}{\fnsymbol{footnote}}

\newcommand{\etap}{\eta^{(}{}'{}^{)}}

\makeatletter
\def\slash#1{\rlap/{\mkern-2mu {#1}}} 
\makeatother

\begin{document}

\begin{titlepage}
\begin{flushright}
\begin{tabular}{l}
IPPP/07/30\\
DCPT/07/60
\end{tabular}
\end{flushright}
\vskip1.5cm
\begin{center}
{\Large \bf \boldmath
$B\to \etap$ Form Factors in QCD}
\vskip1.3cm 
{\sc
Patricia Ball\footnote{Patricia.Ball@durham.ac.uk} and
G.W.~Jones\footnote{G.W.Jones@durham.ac.uk}}
  \vskip0.5cm
        {\em IPPP, Department of Physics,
University of Durham, Durham DH1 3LE, UK} 

\vskip2cm


\vskip6cm

{\large\bf Abstract\\[10pt]} \parbox[t]{\textwidth}{
We calculate the semileptonic form factors $f_+^{B\to \eta}(q^2)$ and
$f_+^{B\to \eta'}(q^2)$ from QCD sum rules on the light-cone (LCSRs), to NLO
in QCD, and for small to moderate $q^2$, $0\leq q^2\leq 16\,{\rm
  GeV}^2$. We include in particular the so-called singlet
contribution, i.e.\ weak annihilation of the $B$ meson with the
emission of two gluons which, thanks to the U(1)$_{\rm A}$ anomaly,
couple directly to $\etap$. This effect is included to leading-twist
accuracy. This contribution has been neglected
in previous calculations of the form factors from LCSRs. We find that 
the singlet contribution to $f_+^{B\to \eta'}$ can be up to $20\%$, while
that to  $f_+^{B\to \eta}$ is, as expected, much smaller and below $3\%$.
We also suggest to measure the ratio ${\cal B}(B\to\eta' e \nu)/{\cal
  B}(B\to \eta e \nu)$ to better constrain the size of the singlet
contribution.} 

\vfill

\end{center}
\end{titlepage}

\setcounter{footnote}{0}
\renewcommand{\thefootnote}{\arabic{footnote}}

\newpage

\section{Introduction}
\setcounter{equation}{0}

$B\to \etap$ transitions are interesting for a number of
reasons: at tree-level, they involve a $b\to u $ transition and hence
are sensitive to the CKM matrix element $|V_{ub}|$. Its precise
determination is crucial for the interpretation of the ``tension'' \cite{UTfit}
that has emerged between the determination of $|V_{ub}|$ from, on the
one hand,
inclusive semileptonic $B\to X_u \ell\nu$ decays \cite{inclusive}, 
and, on the other hand, 
global fits \cite{UTfit,CKMfitter} and the exclusive decay
$B\to\pi\ell\nu$ \cite{Vub,PBvub,FN07a,HPQCD}. The inclusive value of
$|V_{ub}|$ is larger than that from  other determinations
and hints at a non-zero new-physics contribution to the $B_d$
mixing phase $\phi_d$, i.e.\ $\phi_d\neq 2\beta$ \cite{BF06}. While 
an analysis of all available experimental and theoretical information
on $B\to\pi\ell\nu$ found no ``significant'' disagreement between the
exclusive and the inclusive values of $|V_{ub}|$ \cite{FN07a}, the situation
has changed very recently, when the HPQCD lattice collaboration 
reported a mistake in
their calculation of the form factor $f_+^{B\to\pi}$ published in 
Ref.~\cite{HPQCD}; the corrected form factor is larger and hence
yields a smaller
$|V_{ub}|$ \cite{erratum}. The authors of Ref.~\cite{FN07a} have 
since then published
an update \cite{FN07b} of their previous analysis and now conclude that the
exclusive value of  $|V_{ub}|$ is in perfect agreement with the
determination from global fits and that ``the
hints of a disagreement with inclusive determinations of $|V_{ub}|$ are
strengthened''. Also very recently, Neubert has argued
\cite{Neubert} that
the value of  $|V_{ub}|$ obtained by the HFAG collaboration
\cite{HFAG} is dominated by observables with small efficiency and that,
selecting observables with maximum efficiency instead, the resulting
$|V_{ub}|$ is smaller than the HFAG average. Given this
situation it is important to collect information on $|V_{ub}|$ also
from other exclusive processes. $B\to\etap \ell\nu$ decays offer the
opportunity for doing so.

\begin{figure}
$$\epsfxsize=0.4\textwidth\epsffile{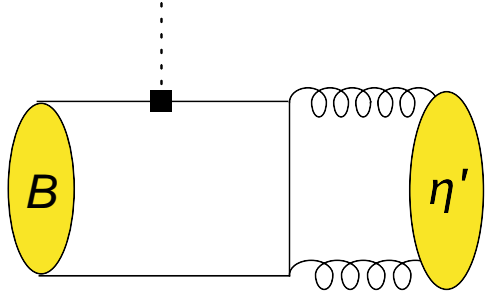}$$
\caption[]{\small Flavour-singlet contribution to a generic $B\to
  \eta'$ transition.}\label{fig1}
\end{figure}

Another reason why $B\to\etap$ transitions are interesting is their
sensitivity to $\eta$-$\eta'$ mixing and the effects of the U(1)$_{\rm
 A}$ anomaly, which is responsible for the large mass of the $\eta'$
and also induces potentially large flavour-singlet contributions to
amplitudes involving $\etap$. 
Indeed the unexpectedly large branching fractions of
inclusive $B\to \eta' X$ and exclusive $B\to \eta' K$ decays, as
compared to e.g.\  $B\to\pi$ transitions, have been attributed to an
enhanced flavour-singlet contribution \cite{anomalous}, which is defined
as the amplitude for producing either a quark-antiquark pair in a singlet
state $(u\bar u + d\bar d + s\bar s)$ which does not contain the $B$'s
spectator quark, or a pair of gluons, followed by hadronization 
into an $\etap$. A generic contribution of this type is
shown in Fig.~\ref{fig1}.
In Ref.~\cite{BN02} it was found that a
rather large singlet-contribution of ca.~$30\%$ to
the form factor $f_+^{B\to\eta'}$ would bring the central values of theoretical
predictions for $B\to \eta' K$ observables in QCD factorisation into
good agreement with experimental results, although the theoretical
uncertainties are too large to allow a definite conclusion on the size
of the singlet contributions. On the other hand, a more recent
analysis of $B$ decays with isosinglet final states, formulated in
SCET, finds that, because of large experimental uncertainties of the
data used to fit non-perturbative parameters, the singlet
contribution to form factors is consistent with 0 \cite{zupan}. 

While the interplay of singlet and octet contributions is 
well understood at the level of local matrix elements, i.e.\ decay
constants (wave functions at the origin) \cite{FKS,TF,EF}, 
less is known about the shape of these wave functions, which are
relevant for dynamical quantities like form factors.
In frameworks based on QCD factorisation 
the mesons' Fock-state wave functions enter in the form of
light-cone distribution amplitudes (DAs).
Constraints on the leading parameters of these DAs
have been obtained from the analysis of the $\etap\gamma$
transition form factor \cite{Kroll97,Kroll,AP03} and  of the
inclusive decay $Y(1S)\to \eta' X$ \cite{AP03}. 
In principle, these DAs can also be constrained from a measurement of the
form factors of $B\to\etap$, for instance from
${\cal B}(B\to\eta'\ell\nu)/{\cal B}(B\to\eta\ell\nu)$, 
as suggested in Ref.~\cite{Kim}. 

Despite the strong phenomenological interest in the size of the
singlet contribution to $f_+^{B\to\etap}$, there is, to the best of
our knowledge, only a single calculation available, based on the
perturbative QCD approach \cite{Li}. Ref.~\cite{Li} finds that this
contribution is negligible in $f_+^{B\to\eta}$, and reaches a few
percent in $f_+^{B\to\eta'}$. Another well-known method for the
calculation of $B\to$~light meson form factors are QCD sum rules on
the light cone (LCSRs) \cite{koles,LCSRs,BZ04}. 
Ref.~\cite{BZ04}, for instance, provides form
factors for $B\to(\pi,K,\eta)$ decays, but does not include the singlet
contribution to $B\to\eta$, nor a calculation of $B\to\eta'$ form
factors. It is the purpose of this paper to remedy this situation and
complete the calculation of $B\to$~light pseudoscalar 
meson form factors from LCSRs by including also the flavour-singlet
contributions. 

Our paper is organised as follows: in Sec.~\ref{sec2} we define the
two most common $\eta$-$\eta'$ mixing schemes and review 
$\etap$ DAs. In Sec.~\ref{sec3} we derive LCSRs 
for the $B\to\etap$ form factors. In Sec.~\ref{sec4} we present
results and conclude.

\section{\boldmath $\eta$ and $\eta'$ Mixing and 
Distribution Amplitudes}\label{sec2}

There are two different mixing schemes in use to describe the
$\eta$-$\eta'$ system: the singlet-octet (SO) and the quark-flavour
scheme (QF)
\cite{FKS}. In the former, the couplings of the relevant 
axial-vector currents to the meson $P=\eta,\eta'$ are given by 
\begin{equation}\label{3}
\langle 0 | J^i_{\mu 5}| P(p)\rangle = i f_P^i p_\mu \quad (i=1,8),
\end{equation}
where $J_{\mu 5}^8$ denotes the SU(3)$_{\rm F}$-octet and $J_{\mu
  5}^1$ the  SU(3)$_{\rm F}$-singlet axial-vector current,
  respectively. The four parameters $f_P^i$ define two decay constants
  $f_i$ of a hypothetical pure singlet or octet state $|\eta_i\rangle$
  and also two mixing angles $\theta_i$ via
\begin{equation}
\left(
\begin{array}{cc}
f_\eta ^8 & f_\eta^1 \\
f_{\eta'} ^8 & f_{\eta'}^1 
\end{array}\right)
= 
\left(
\begin{array}{cc}
\cos\theta_8 & -\sin\theta_1 \\
\sin\theta_8 & \phantom{-}\cos\theta_1 
\end{array}\right)
\left(
\begin{array}{cc}
f_8 & 0 \\
0 & f_1
\end{array}
\right).
\label{corr}
\end{equation}
The advantage of this scheme is that the impact of the U(1)$_{\rm A}$
anomaly is plainly localised in $f_1$, via the divergence of the
singlet current $J_{\mu 5}^1$, while $\theta_i\neq 0$
and $f_8\neq f_\pi$ are SU(3)$_{\rm F}$-breaking effects. By the same
token, the SO scheme also diagonalises the
renormalisation-scale
dependence of parameters and hence is very useful for checking the
cancellation of divergences in perturbative calculations: 
$f_8$ and $\theta_i$ are
scale-independent, while $f_1$ renormalises multiplicatively \cite{leut}:
\begin{equation}\label{scaledep}
\mu\,\frac{d f_1}{d \mu} = - n_f \left(\frac{\alpha_s}{\pi}\right)^2
f_1 + O(\alpha_s^3)\,.
\end{equation}
In the QF mixing scheme, on the other hand, the basic axial-vector currents are
\begin{equation}
J^q_{\mu 5} = \frac{1}{\sqrt{2}} \left(\bar u \gamma_\mu \gamma_5 u + 
\bar d \gamma_\mu \gamma_5 d \right),\qquad
J^s_{\mu 5} = \bar s \gamma_\mu \gamma_5 s\,,
\end{equation}
and the corresponding couplings to $P=\eta,\eta'$ are given by
\begin{equation}\label{6}
\langle 0 | J^r_{\mu 5}| P(p)\rangle = i f_P^r p_\mu \quad (r=q,s)\,.
\end{equation}
In complete correspondence to (\ref{corr}) one  has
\begin{equation}
\left(
\begin{array}{cc}
f_\eta ^q & f_\eta^s \\
f_{\eta'} ^q & f_{\eta'}^s 
\end{array}\right)
= 
\left(
\begin{array}{cc}
\cos\phi_q & -\sin\phi_s \\
\sin\phi_q & \phantom{-}\cos\phi_s 
\end{array}\right)
\left(
\begin{array}{cc}
f_q & 0 \\
0 & f_s
\end{array}
\right).
\label{corr2}
\end{equation}
The basic difference to the SO scheme is that now the difference
between the two angles $\phi_{q,s}$ is not caused by SU(3)$_{\rm F}$
effects, like that between $\theta_1$ and $\theta_8$, but by an
OZI-rule violating contribution, as explained in Ref.~\cite{TF}. While the
numerical values of $\theta_i$ differ largely, with typical values
$\theta_8\approx -20^\circ$ and $\theta_1\approx - 5^\circ$, one finds
$\phi_s-\phi_q\, \raisebox{-3pt}{$\stackrel{<}{\sim}$}\, 
5^\circ$, with $\phi_q\approx
\phi_s \approx 40^\circ$ \cite{FKS,TF}. This led the authors of
Ref.~\cite{FKS} to suggest the QF scheme as an approximation to
describe $\eta$-$\eta'$ mixing, based on neglecting the difference
$\phi_q-\phi_s$ (and all other OZI-breaking effects):
\begin{equation}
\phi\equiv \phi_{q,s},\qquad \phi_q-\phi_s\equiv 0\,.
\end{equation}
The advantage of this scheme is that it has only 3 parameters, $f_q$,
$f_s$ and $\phi$, which implies that the mixing of states is the same
as that of the decay constants:
\begin{equation}\label{8}
\left(
\begin{array}{c}
\eta \\ \eta'
\end{array}
\right) 
= 
\left(
\begin{array}{ll}
\cos\phi & -\sin\phi\\
\sin\phi & \phantom{-}\cos\phi
\end{array}
\right)
\left(
\begin{array}{c}
\eta_q \\ \eta_s
\end{array}
\right) .
\end{equation}
The disadvantage is that, due to the neglection of OZI-breaking effects, 
the re\-nor\-ma\-li\-sa\-tion-scale dependence of $f_1$ is not reproduced 
-- as it is induced precisely by OZI-breaking terms
\cite{TF}. While this is not really an issue numerically, as the
scale-dependence of $f_1$ is a two-loop effect, Eq.~(\ref{scaledep}),
the problem of the incompatibility of the QF scheme with the
scale-dependence of parameters 
will come back at the level of non-local matrix elements,
i.e.\ DAs, see below.

Given enough data to fix all independent parameters, there is no reason
to prefer the QF over the SO scheme. For DAs, however, the SO scheme
leads to a proliferation of unknown parameters, while the QF scheme is
more restrictive, see below. For this reason we decide
to use the QF scheme in this paper. Its basic parameters have been
determined as \cite{FKS}
\begin{equation}
f_q  =  (1.07\pm 0.02)f_\pi,\qquad f_s = (1.34\pm
0.06)f_\pi\,,\qquad
\phi =  39.3^\circ\pm 1.0^\circ\,.
\end{equation}
This can be translated into values for the SO parameters as
\begin{eqnarray}
f_8 & = & \sqrt{\frac{1}{3}\,f_q^2 + \frac{2}{3} f_s^2} = (1.26\pm
0.04) f_\pi\,,\nonumber\\
f_1  &=&  \sqrt{\frac{2}{3}\,f_q^2 + \frac{1}{3} f_s^2} = (1.17\pm
0.03) f_\pi\,,\nonumber\\
\theta_8 & = & \phi-{\rm arctan}[\sqrt{2} f_s/f_q] = -21.2^\circ \pm
1.6^\circ\,,\nonumber\\
\theta_1  &=&  \phi-{\rm arctan}[\sqrt{2} f_q/f_s] = -9.2^\circ \pm
1.7^\circ\,.
\end{eqnarray}
Note that in the QF scheme $f_{q,s}$ are scale-independent parameters,
and so is $f_1$ as obtained from the above relations.
The SO decay constants can be expressed in terms of the QF
ones and the angle $\phi$ as
\begin{equation}\label{11}
\left(
\begin{array}{cc}
f_\eta ^8 & f_\eta^1 \\
f_{\eta'} ^8 & f_{\eta'}^1 
\end{array}\right)
= 
\left(
\begin{array}{cc}
\cos\phi & -\sin\phi \\
\sin\phi & \phantom{-}\cos\phi 
\end{array}\right)
\left(
\begin{array}{cc}
f_q & 0 \\
0 & f_s
\end{array}\right)
\left(
\begin{array}{cc}
\phantom{-}\sqrt{\frac{1}{3}} & \sqrt{\frac{2}{3}}\\
-\sqrt{\frac{2}{3}} & \sqrt{\frac{1}{3}}
\end{array}\right).
\end{equation}

Let us now turn to light-cone DAs, that is the extension of matrix
elements like (\ref{3}) and (\ref{6}) to those over non-local operators on the
light-cone. This paper is not the place to give a thorough discussion
of the properties of DAs, for which we refer to reviews
\cite{CZ} and to Refs.~\cite{PB98,BBL06}. Suffice it to say that the DAs
are ordered in terms of increasing twist, with the minimum, or
leading, twist for meson DAs being two. Motivated by the structure of
the evolution of DAs under a change of the renormalisation scale
$\mu$, they are expanded in terms of so-called asymptotic DAs
multiplied by Gegenbauer polynomials. In the context of this paper it
is important to recall that the U(1)$_{\rm A}$ anomaly induces, in
addition to two-quark DAs, also two-gluon DAs, of both leading and higher
twist. Some properties of these higher-twist DAs have been studied in
Ref.~\cite{AP03}. In this paper we only include the effects of the
leading-twist two-gluon DA, which is justified as its effects turn out
to be small and higher-twist DAs are estimated to have even smaller impact.
We will come back to that in Sec.~\ref{sec4}.

We define the twist-2 two-quark DAs of $\etap$ as \cite{Kroll}
\begin{equation}
\langle 0 | \bar\Psi(z) {\cal C}_i \slash{z} \gamma_5 [z,-z] \Psi(-z)
| P(p)\rangle = i (pz) f_P^i \int_0^1 du\, e^{i \xi (pz)}
\phi_{2;P}^i(u) \,.
\end{equation}
Here $z_\mu$ is a light-like vector, $z^2=0$, and
$[x,y]$ stands for the path-ordered gauge factor along the straight line
connecting the points $x$ and $y$,
\begin{equation}
[x,y] ={\rm P}\exp\left[ig\!\!\int_0^1\!\! dt\,(x-y)_\mu
  A^\mu(tx+(1-t)y)\right].
\label{Pexp}
\end{equation}
$u$ ($1-u$) is the momentum fraction carried by the quark (antiquark)
in the meson,
$\xi$ is short for $2u-1$. $\phi_{2;P}^i(u)$ is the twist-2 DA of the meson
$P$ with respect to the current whose flavour content is given 
by ${\cal C}_i$, with $\Psi = (u,d,s)$ the triplet of light-quark fields in
flavour space. For the SO currents, one has ${\cal
  C}_1 = \mbox{\boldmath $1$}/\sqrt{3}$ and ${\cal C}_8 =
\lambda_8/\sqrt{2}$, while for the QF currents ${\cal C}_q =
(\sqrt{2} {\cal C}_1 + {\cal C}_8)/\sqrt{3}$ and 
${\cal C}_s = ({\cal C}_1 - \sqrt{2} {\cal
  C}_8)/\sqrt{3}$, with $\lambda_i$ the standard Gell-Mann matrices.

The gluonic twist-2 DA is defined as\footnote{This
  definition refers to the ``$\sigma$-rescaled'' DA
  $\phi^\sigma_g$ in Ref.~\cite{Kroll} with 
$\sigma = \sqrt{3}/C_F$. It
  agrees with that used in Refs.~\cite{AP03,Li}, which means that we can
  use their results for the two-gluon Gegenbauer moment
  $B^g_2$ without rescaling.} 
\begin{equation}
\langle 0 | G_{\mu z}(z) [z,-z] \widetilde G^{\mu z}(-z) | P(p)\rangle
= \frac{1}{2}\,(pz)^2 \frac{C_F}{\sqrt{3}} f_P^1 \int_0^1 du\, e^{i\xi (pz)}
\psi_{2;P}^g(u)\,.
\end{equation}
In order to perform the calculation of the correlation function
defined in the next section, we also need the matrix element of the
meson $P$ over two gluon fields. Dropping the gauge factor $[z,-z]$,
one has
\begin{equation}\label{15}
\langle 0 | A^A_\alpha(z) A^B_\beta(-z) | P(p)\rangle =
\frac{1}{4}\,\epsilon_{\alpha\beta\rho\sigma} \,\frac{z^\rho
  p^\sigma}{(pz)} \,\frac{C_F}{\sqrt{3}}\, f_P^1 \,\frac{\delta^{AB}}{8}
\int_0^1 du\, e^{i\xi (pz)}\,\frac{\psi_{2;P}^g(u)}{u(1-u)}\,.
\end{equation}

Because of the positive G-parity of $\eta$ and $\eta'$, 
the two-quark DAs are symmetric under
$u\leftrightarrow 1-u$:
\begin{equation}
\phi_{2;P}^i(u) = \phi_{2;P}^i(1-u)\,;
\end{equation}
they are expanded in terms of Gegenbauer polynomials as 
\begin{equation}
\phi_{2;P}^i(u) = 6 u (1-u) \left( 1 + \sum_{n=2,4,\dots} a_n^{P,i}(\mu)
C^{3/2}_n(\xi) \right) \quad (i=1,8,q,s)\,;
\end{equation}
$a_n^{P,i}$ are the quark Gegenbauer moments.
As for the two-gluon DAs, the asymptotic DA is
$u^{2j-1}(1-u)^{2j-1}$ with $j=3/2$ the lowest conformal spin of the
operator $G_{\mu z}$; the expansion goes in terms of Gegenbauer
polynomials $C^{5/2}_n$. One can show that $\psi_{2;P}^g$ is
antisymmetric:
\begin{equation}
\psi_{2;P}^g(u) = - \psi_{2:P}^g(1-u)\,;
\end{equation}
in particular $\int_0^1 du\, \psi_{2;P}^g(u) = 0$ and the local
twist-2 matrix element $\langle 0 |
G_{\mu z} \widetilde G^{\mu z}| P\rangle$ vanishes. The non-vanishing
coupling $\langle 0 | G_{\alpha\beta}  \widetilde G^{\alpha\beta}|
P\rangle$ induced by the U(1)$_{\rm A}$ anomaly is a twist-4
effect. The corresponding matrix elements are given, in the QF
scheme, by \cite{FKS}:
\begin{eqnarray}
\langle 0 | \alpha_s G\tilde{G}/(4\pi) | \eta_q\rangle & = & 
f_s (m_\eta^2-m_{\eta'}^2) \sin\phi \cos\phi\,,\nonumber\\
\langle 0 | \alpha_s G\tilde{G}/(4\pi) | \eta_s\rangle & = & 
f_q (m_\eta^2-m_{\eta'}^2)/\sqrt{2} \sin\phi \cos\phi\,.\label{extra}
\end{eqnarray}
We will estimate the size of these effects in Sec.~\ref{sec4}. There
are no twist-3 two-gluon DAs and the remaining twist-4 DAs also have
vanising normalisation, see Ref.~\cite{AP03}. 
The conformal expansion of the twist-2 two-gluon DA reads
\begin{equation}
\psi_{2;P}^g(u,\mu) = u^2 (1-u)^2 \sum_{n=2,4,\dots} B^{P,g}_n(\mu)
C^{5/2}_{n-1}(\xi)
\end{equation}
with the gluonic Gegenbauer moments $B^{P,g}_n$. In this paper, we
truncate both $\phi^i_{2;P}$ and $\psi^g_{2;P}$ at $n=2$. This is
due to the fact that our knowledge about these higher-order
Gegenbauer moments is very restricted. An estimate of the effect of
higher Gegenbauer moments in $\phi_{2;\pi}$ on the $B\to\pi$ form
factor $f_+^\pi$ has been given in Ref.~\cite{angi}, based on a
certain class of models for the full DA beyond conformal
expansion. The effect of neglecting $a_{n\geq 4}^\pi$ we found to be
very small, $\sim 2\%$. We expect the truncation error from neglecing
$B^g_{n\geq 4}$ to be of similar size.

\begin{figure}
$$\epsfxsize=0.45\textwidth\epsffile{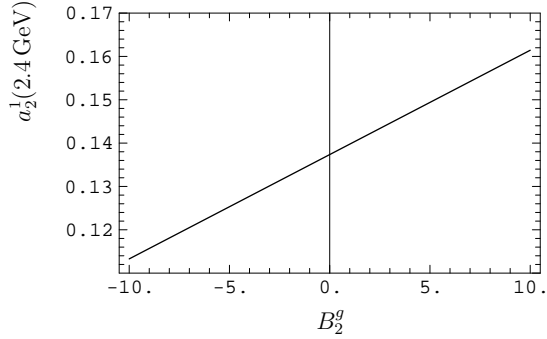}$$
\vskip-15pt
\caption[]{\small Dependence of $a_2^1(2.4\,{\rm GeV})$ on $B_2^g(1\,{\rm
    GeV})$, Eq.~(\ref{21}), for $a_2^1(1\,{\rm GeV}) = 0.2$.}\label{fig2}
\end{figure}

$\phi_{2;P}^1$ and $\psi_{2;P}^g$ mix upon evolution in $\mu$, 
see for instance Ref.~\cite{Kroll}.  This amounts to a mixing of
$a_2^{P,1}$ and $B^{P,g}_2$, resulting in the renormalisation-group
equation, to LO accuracy,
\begin{equation}\label{20}
\mu\,\frac{d}{d\mu}\left(\begin{array}{c} a_2^1\\ B^g_2
\end{array}\right)
=
-\frac{\alpha_s}{4\pi} \left(\begin{array}{cc} \displaystyle\frac{100}{9} &
\displaystyle  -\frac{10}{81}\\\vphantom{\displaystyle\frac{100}{9}}
 -36 & 22\end{array}\right)
\left(\begin{array}{c} a_2^1\\ B^g_2
\end{array}\right),
\end{equation}
where for simplicity we have dropped the superscript $P$. We only
quote the solution for $a_2^1$:
\begin{eqnarray}
a_2^1(\mu) & = & \left[ \left(\frac{1}{2} -
  \frac{49}{2\sqrt{2761}}\right) L^{\gamma_2^+/(2\beta_0)} + 
\left(\frac{1}{2} +
  \frac{49}{2\sqrt{2761}}\right) L^{\gamma_2^-/(2\beta_0)}\right]
  a_2^1(\mu_0) \nonumber\\
&&{}+ \frac{5}{9\sqrt{2761}}\left[L^{\gamma_2^-/(2\beta_0)}-
L^{\gamma_2^+/(2\beta_0)}\right] B_2^g(\mu_0)\label{21}
\end{eqnarray}
with $L = \alpha_s(\mu)/\alpha_s(\mu_0)$ and the anomalous dimensions
$\gamma_2^\pm = (149\pm \sqrt{2761})/9$. This is to be compared to the
evolution of the octet Gegenbauer moment:
\begin{equation}\label{22}
a_2^8(\mu) = L^{50/(9\beta_0)} a_2^8(\mu_0)\,.
\end{equation}
Numerically, the evolution of $a_2^1$ does not differ much from that
of $a_2^8$, for a wide range of $B_2^g$: assume $a_2^8(1\,{\rm GeV})
\equiv a_2^1(1\,{\rm GeV})$, as is the case for a strict imposition of
the QF scheme. Choose $a_2^8(1\,{\rm GeV})=0.2$, as indicated by our
knowledge of twist-2 DAs of the $\pi$; then we have $a_2^8(2.4\,{\rm
  GeV}) = 0.137$ from (\ref{22}); $2.4\,$GeV is 
a typical scale in the calculation of form factors
from LCSRs. In Fig.~\ref{fig2} we show the results of the evolution of
the singlet Gegenbauer moment $a_2^1$ from 1 to 2.4$\,$GeV, from 
Eq.~(\ref{21}), for the
 range of gluon Gegenbauer moments $|B_2^g(1\,{\rm
  GeV})|<10$. Evidently the impact of the different anomalous
dimensions of $a_2^1$ and $a_2^8$ is  negligible
($a_2^1(2.4\,{\rm GeV}) = 0.137$ for $B^g_2=0$) and the mixing of
$B_2^g$ into $a_2^1$ is smaller than 20\% within the range of 
$B_2^g$ considered.

At this point we would like to come back to the impact of evolution on
the consistency of the QF scheme. 
We introduce the twist-2 two-quark DAs $\phi_2^i$, $i=1,8,q,s$,
corresponding to the basis states $|\eta_i\rangle$ in the SO and QF
scheme, respectively. We then have, in terms of the quark
valence Fock states $|q\bar q\rangle$ and $|s\bar s\rangle$ \cite{Kroll}: 
\begin{equation}
|\eta_q\rangle \sim \phi_2^q (u) |q\bar q\rangle + \phi_2^{\rm
 OZI}(u) |s\bar s\rangle\,,\quad
|\eta_s\rangle \sim  \phi_2^{\rm OZI} (u) |q\bar q\rangle + 
\phi_2^s(u) |s\bar s\rangle\,,
\end{equation}
where $q\bar q$ is shorthand for $(u\bar u + d\bar d)/\sqrt{2}$ and 
\begin{equation}
\phi_2^q = \frac{1}{3}\,(\phi_2^8 + 2\phi_2^1)\,,\quad
\phi_2^s = \frac{1}{3}\,(2\phi_2^8 + \phi_2^1)\,,\quad
\phi_2^{\rm OZI} = \frac{\sqrt{2}}{3} (\phi_2^1-\phi_2^8)\,.\label{24}
\end{equation}
In the QF scheme, the ``wrong-flavour'' DA 
$\phi_2^{\rm OZI}$, which is generated by OZI-violating interactions, 
is set to 0. Once this is done at
a certain scale, however, the different evolution of $a_n^1$ and
$a_n^8$, Eqs.~(\ref{21}) and (\ref{22}), 
will generate a non-zero $\phi_2^{\rm OZI}$ already to LO
accuracy. A consistent implementation of the QF scheme hence requires
one to either set $a_n^{1,8}\equiv 0$ and also $B^g_n\equiv 0$, or to
set $a_n^8\equiv a_n^1$ and neglect the different 
scale-dependence of these parameters. In practice, however, the QF scheme is an
approximation anyway, motivated by the observed smallness of one parameter, the
difference of mixing angles $\phi_s-\phi_q$. The induced non-zero
DA $\phi_2^{\rm OZI}$ is numerically very small for the scales relevant
for our calculation, $\mu=1\,$GeV and $2.4\,$GeV. We hence implement
the QF scheme for DAs as follows: we set $\phi_2^1\equiv \phi_2^8$ at
the scale $\mu=1\,$GeV, which, by virtue of (\ref{24}), implies
$\phi_2^q\equiv\phi_2^s$ at the same scale. We then evolve $a_2$
according to the scaling-law for the octet Gegenbauer moment,
Eq.~(\ref{22}).\footnote{This is equivalent to imposing the QF-scheme
  relation $a_2^1=a_2^8$ as the scale $\mu=2.4\,$GeV and defining
  $B^g_2$ as $B^g_2(2.4\,{\rm GeV})$.} 
We also set $\psi_{2;\eta}^g=\psi_{2;\eta'}^g$; again
any SU(3)$_{\rm F}$ breaking of this relation is expected to have
only very small impact on $f_+^{B\to\etap}$. The twist-2 parameters
used in our calculation are then reduced to 2: $a_2$ and $B^g_2$. For
error estimates, we will also sometimes distinguish between $a_2^\eta$
and $a_2^{\eta'}$.

As far as numerics is concerned, we assume that the bulk of 
SU(3)$_{\rm F}$-breaking effects is described by the decay constants
via $f_q\neq
f_\pi$, and that SU(3)$_{\rm F}$ breaking in Gegenbauer moments is
subleading. This motivates setting $a_2^q = a_2^\pi$, 
with $a_2^\pi(1\,{\rm GeV}) = 0.25\pm
0.15$ as an average over a large number of 
calculations and fits to experimental data \cite{BBL06}; this number
also agrees with a recent lattice determination \cite{a2pilatt}.
$a_2^q=a_2^\pi$ is justified as, as 
discussed in Ref.~\cite{BBL06}, there is no evidence for noticeable
SU(3)-breaking effects between $a_2^\pi$ and $a_2^K$ and the main
SU(3)-breaking in the DAs is due to non-zero odd Gegenbauer
moments. In this work we only need $a_2^q$, and as a QCD sum rule for
this parameter would look essentially the same as that for $a_2^\pi$,
except for a slightly 
different value for the decay constant, $f_\pi\neq f_q$, and different
numerical values for the continuum threshold $s_0$ and the window in
the Borel parameter $M^2$, we see no plausible source for large SU(3)
breaking between $a_2^\pi$ and $a_2^q$. 
To the best of our knowledge, no calculation of $B_2^g$ is available. 
Results from
fits to data have been obtained from the $\etap\gamma$ transition form
factor, yielding $B_2^g(1\,{\rm GeV}) = 9\pm 12$ \cite{Kroll}, and the
combined analysis of this form factor and the inclusive decay $Y(1S)\to
\eta' X$ yielding $B_2^g(1.4\,{\rm GeV}) = 4.6\pm 2.5$ \cite{AP03}. 
These results, however, have to be taken cum grano salis
as they are highly correlated with the
simultaneous determination of $a_2^1$ and $a_2^8$ from the same data,
yielding $a_2^1(1\,{\rm GeV}) = -0.08\pm 0.04$, $a_2^8(1\,{\rm GeV}) =
-0.04\pm 0.04$ \cite{Kroll}  and $a_2^1(1.4\,{\rm GeV}) = a_2^8(1.4\,{\rm GeV})
= -0.054\pm 0.029$ \cite{AP03}. The same analysis applied to the
$\pi\gamma$ form factor returns $a_2^\pi (1\,{\rm GeV}) =-0.06\pm
0.03$ \cite{vogt}. These results are not really compatible with those
from the direct calculation of $a_2^\pi$ from lattice and QCD
sum rules; in particular the sign of $a_2^\pi$ is unambiguously fixed
as being positive. A possible reason for this discrepancy is the neglect of
higher-order terms in the light-cone expansion and that, in addition, as
one of the photons in the process is nearly real with virtuality
$q^2\approx 0$, one also has to take into account long-distance
photon interactions, of order $1/\sqrt{q^2}$ \cite{rady}.
For this reason, we assume the very conservative range
$B_2^g(2.4\,{\rm GeV}) = 0\pm
20$ in the remainder of this paper.

As far as higher-twist DAs are concerned, we only need
those involving currents with flavour content $\bar q q = (\bar u u +
\bar d d)/\sqrt{2}$. In line with the implementation of the QF scheme
for twist-2 DAs, we include SU(3)$_{\rm F}$ breaking only via
the decay constants and set
\begin{eqnarray}
\frac{1}{f^q_{\etap}}\,\langle 0 | \bar \Psi(z) C_q [z,-z] \Gamma \Psi(-z) |
\etap(p) \rangle & = & \frac{1}{f_\pi}\,
\langle 0 | \bar d(z)[z,-z]\Gamma u(-z) |
\pi^-(p) \rangle\,,\nonumber\\
\frac{1}{f^q_{\etap}}\,\langle 0 | \bar \Psi(z)[z,vz] G(vz) C_q \Gamma
     [vz,-z] \Psi(-z) |
\etap(p) \rangle 
& = &\nonumber\\ 
\lefteqn{
\frac{1}{f_\pi}\,\langle 0 | \bar d(z)[z,vz] G(vz)\Gamma [vz,-z] u(-z) |
\pi^-(p) \rangle\,,}\hspace*{3cm}
\end{eqnarray}
where $\Gamma$ is the relevant Dirac structure and $G(vz)$ the gluon
field-strength tensor. The precise definitions of all twist-3 and 4
DAs, as well as up-to-date numerical values of the $\pi$'s hadronic parameters
can be found in Ref.~\cite{BBL06}.
Let us shortly comment on the validity of this treatment for twist-3
two-quark DAs. As is well known, the normalisation of these DAs is
given, for the $\pi$, by $f_\pi m_\pi^2/(2m_q)$ and enters the
light-cone sum rules for $B\to\pi$ transitions as a $1/m_b$
correction, see explicit formulas for the corresponding $D$ form
factor in Ref.~\cite{Ds}. Although
suppressed by one power of the heavy quark mass, this contribution is
numerically non-negligible due to the chiral enhancement factor. Following the
above implementation of SU(3) breaking, we set $f_\pi m_\pi^2/(2m_q)
\to f_q m_\pi^2/(2m_q)$ for $\eta_q$ (the corresponding quantity for
$\eta_s$ is not needed). In contrast, the 
inclusion of all SU(3) effects leads one
to consider the quantity
\begin{equation}
h_q = f_q (m_\eta^2 \cos^2\phi + m_{\eta'}^2 \sin^2\phi) - \sqrt{2}
f_s (m_{\eta'}^2 - m_\eta^2) \sin\phi \cos\phi\,;\label{argh}
\end{equation}
the normalisation of the twist-3 DAs of $\eta_q$ is given by $h_q/(2m_q)$.
To leading order in the chiral expansion and $1/N_c$ expansion, 
$h_q\to f_q m_\pi^2 =
0.0025\,{\rm GeV}^3$, which
is the value used in our scheme. As discussed in Ref.~\cite{BN02}, the
full expression (\ref{argh}) yields $h_q = (0.0015\pm 0.004)\,{\rm
  GeV}^3$, i.e.\ a 200\% uncertainty, if the errors of $f_{q,s}$ and
$\phi$ are treated as uncorrelated. The large error is due to a
cancellation between the two terms in (\ref{argh}). As the parameter
we need is actually $h_q/(2m_q)$, with $m_q$ not very well constrained
(yet) from lattice calculations\footnote{A recent unquenched
  calculation yields $\overline{m} \equiv (m_u+m_d)/2 =
  (3.54^{+0.64}_{-0.35})\,$MeV at the scale $\mu=2\,$GeV 
\cite{lattmq}.} and the correlation of the errors of $f_{q,s}$ and
$\phi$ is not known, we feel that a total 250\% uncertainty of
$h_q/(2m_q)$ is slightly exaggerated and an artifact of the numerical
cancellation. Instead, we work to leading order in the chiral
expansion and set 
$h_q/(2m_q) = f_q B_0$, with $B_0
= m_\pi^2/(2m_q) = -2 \langle 0 | \bar q q | 0 \rangle/f_\pi^2$
\cite{PB98}. $\langle 0 | \bar q q | 0 \rangle$, the quark condensate,
is the order parameter of chiral symmetry breaking and known from QCD
sum rules to have the value $\langle 0 | \bar q q | 0 \rangle =
(-0.24\pm 0.01)^3\,{\rm GeV}^3$. From this, one finds $B_0 = (1.6\pm
0.2)\,$GeV \cite{PB98}, 
which, together with the error on $f_q$, implies a total
15\% uncertainty for the normalisation of the twist-3 DAs. This is the
standard treatment of these terms in the framework of light-cone sum rules.

\section{LCSRs for Gluonic Contributions}\label{sec3}

The key idea of light-cone sum rules is to
consider a correlation function of the weak current and a current with
the quantum numbers of the $B$ meson, sandwiched between the vacuum
and an $\eta$ or $\eta'$ state. 
For large (negative) virtualities of these currents, the
correlation function is, in coordinate-space, dominated by distances
close to the light-cone and can be discussed in the framework of
light-cone expansion. In contrast to the short-distance expansion
employed by conventional QCD sum rules \`a la SVZ \cite{SVZ}, where
non-perturbative effects are encoded in vacuum expectation values 
of local operators with
vacuum quantum numbers, the condensates, LCSRs
rely on the factorisation of the underlying correlation function into
genuinely non-perturbative and universal hadron DAs
$\phi$. The DAs are convoluted with process-dependent amplitudes $T_H$,
which are the analogues of the Wilson coefficients in the
short-distance expansion and can be
calculated in perturbation theory. Schematically, one has
\begin{equation}\label{eq:schemat}
\mbox{correlation function~}\sim \sum_n T_H^{(n)}\otimes \phi_n.
\end{equation}
The expansion is ordered in terms of contributions of
increasing twist $n$. 
The light-cone expansion is
matched to the description of the correlation function in terms of hadrons
by analytic continuation
into the physical regime and the application of a Borel
transformation, which introduces the Borel parameter $M^2$ and
exponentially suppresses contributions from higher-mass states.
In order to extract the contribution
of the $B$ meson, one describes the contribution of other hadron states by
a continuum model, which introduces a second model parameter, 
the continuum threshold $s_0$. The sum rule then yields 
the form factor in question, $f_+$, multiplied by the coupling of the
 $B$ meson to
its interpolating field, i.e.\ the $B$ meson's leptonic decay constant
$f_B$.

LCSRs are available for the $B\to\pi,K$ form
factor $f_+$ to
$O(\alpha_s)$ accuracy for the twist-2 and part of the twist-3
contributions and at
tree-level for higher-twist (3 and 4) contributions
\cite{BZ04}. 

We define the $B\to P$ form factors as
\begin{equation}\label{FF}
\langle P(p) | \bar u \gamma_\mu b | B(p+q)\rangle = \left\{
(2p+q)_\mu - \frac{m_B^2-m_P^2}{q^2}\,q_\mu\right\}
\frac{f_+^P(q^2)}{\sqrt{2}} +
\frac{m_B^2-m_P^2}{q^2}\,q_\mu\,\frac{f_0^P(q^2)}{\sqrt{2}}\,.
\end{equation}
Note that we include a factor $1/\sqrt{2}$ on the right-hand side. This
is to ensure that in the limit of SU(3)$_{\rm F}$ symmetry and no
$\eta$-$\eta'$ mixing $f_+^\eta = f_+^\pi$. 

In the semileptonic decay $B \to \eta^{(')} l\nu_l$ the form factor
 $f^P_0$ ($P=\eta,\eta'$) enters proportional to the
lepton mass $m_l^2$ and hence is irrelevant for light leptons 
($l=e,\mu$), where only $f^P_+$ matters.
The semileptonic decay can be used to determine the size of 
the CKM matrix element $|V_{ub}|$ from the spectrum
\begin{equation}\label{eq:spectrum}
\frac{d\Gamma}{dq^2}(B \to \eta^{(')} l \nu_l) = \frac{G_F^2 |V_{ub}|^2}{
192\pi^3m_B^3}\lambda^{3/2}(q^2) |f^P_+(q^2)|^2 \,,
\end{equation}
where $\lambda(x) = (m_B^2+m_{P}^2-x)^2-4m_B^2m_{P}^2$. Alternatively,
as we shall see, the ratio of branching ratios ${\cal B}(B\to\eta'
\ell\nu)/{\cal B}(B\to \eta\ell\nu)$ can be used to constrain the
gluonic Gegenbauer moment $B_2^g$.

Our starting point for calculating $f_{+}^P$ 
is the correlation function
\begin{eqnarray}
\label{eq:corr}
\Pi^P_{\mu}(p,q) &=& i\int d^4x\,e^{i (q x)} \langle P(p)| 
T [\bar u \gamma_\mu b](x) j_B^{\dagger}(0) |0 \rangle\\
&=& \Pi_+^P(q^2,p_B^2) (2p+q)_\mu + \dots\nonumber
\end{eqnarray}
where $j_B= m_b \bar u i\gamma_5 b$ is the interpolating field for the
$B$ meson and $p_B^2=(p+q)^2$ its virtuality. For
\begin{equation}\label{eq:virt}
m_b^2-p_B^2 \geq O(\Lambda_{\rm QCD}m_b), \qquad
m_b^2-q^2 \geq O(\Lambda_{\rm QCD}m_b),
\end{equation}
the correlation function (\ref{eq:corr})
is dominated by light-like distances
and therefore accessible to an expansion around the light-cone. 
The above conditions can be understood by demanding that the exponential 
factor in (\ref{eq:corr}) vary only slowly. The light-cone expansion
is performed by integrating out the transverse and ``minus'' degrees
of freedom and leaving only the longitudinal momenta of the partons as
relevant degrees of freedom. The integration over transverse momenta
is done up to a cutoff, $\mu_{\rm IR}$, all momenta below which are
included in a the DAs $\phi_n$. 
Larger transverse momenta are calculated in perturbation theory. The
correlation function is hence decomposed, or factorised, into
perturbative contributions $T$ and nonperturbative contributions
$\phi$, which both depend on the longitudinal parton momenta and the
factorisation scale $\mu_{\rm IR}$. 
The schematic relation (\ref{eq:schemat}) can then be written 
in more explicit form, including only two-particle DAs, as
\begin{equation}
\label{eq:lcexp}
\Pi_+^P(q^2,p_B^2) = \sum_n \int_0^1 du \,
T^{(n)}(u,q^2,p_B^2,\mu_{\rm{IR}}) \phi_{n;P}(u,\mu_{\rm{IR}}).
\end{equation}
As $\Pi_+$ itself is independent of the arbitrary scale
$\mu_{\rm{IR}}$, the scale-dependence of $T^{(n)}$ and $\phi_{n}$ 
must cancel
each other. If there is more than one contribution of a
  given twist, they will mix under a change of
$\mu_{\rm IR}$ and it is only in the sum of all such contributions
  that the residual $\mu_{\rm IR}$ dependence cancels. This is
   what happens with the two-quark and two-gluon
  contributions to $B\to \etap$.
Eq.~(\ref{eq:lcexp}) is called a ``collinear'' factorisation formula,
as the momenta of the partons in $P$ are collinear with the
$P$'s momentum. Its validity actually has to be verified, which is
done precisely by checking that the $\mu_{\rm IR}$ dependence  cancels. 
In Ref.~\cite{BZ04} it has been shown that the above formula holds to
$O(\alpha_s)$ accuracy for two-quark twist-2 and -3 contributions.

In calculating the correlation function, we use  relation (\ref{8}) between
$|\eta^{(')}\rangle$ and the QF basis states  $|\eta_{q,s}\rangle$,
so that
\begin{equation}\label{32}
\Pi^\eta_\mu = \frac{1}{\sqrt{2}}\left(\Pi^q_{\mu} \cos\phi - \Pi^s_{\mu}
\sin\phi\right),\quad
\Pi^{\eta'}_\mu = \frac{1}{\sqrt{2}}\left(\Pi^q_{\mu} \sin\phi + \Pi^s_{\mu}
\cos\phi\right).
\end{equation}
As the correlation function involves the current $\bar u \gamma_\mu
b$, $\Pi^s_\mu$ vanishes to leading order in $\alpha_s$ and at
$O(\alpha_s)$ is due only to gluonic Fock states of the
meson. $\Pi^q_\mu$, on the
other hand, receives contributions from both quark and gluon
states. The quark contributions have been calculated in
Ref.~\cite{BZ04} for $B\to\pi$, including $O(\alpha_s)$ corrections to
twist-2 and -3 contributions, and to tree-level accuracy for twist-4
contributions. The
corresponding expressions yield
$\Pi^q_+$, with the replacement $f_\pi\to f_q$. 

\begin{figure}
$$
\raisebox{13pt}{\epsfxsize=0.25\textwidth\epsffile{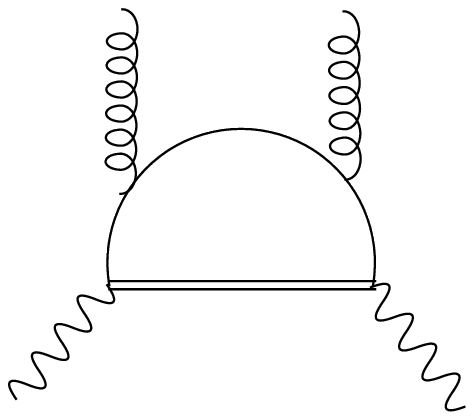}}\qquad
\epsfxsize=0.25\textwidth\epsffile{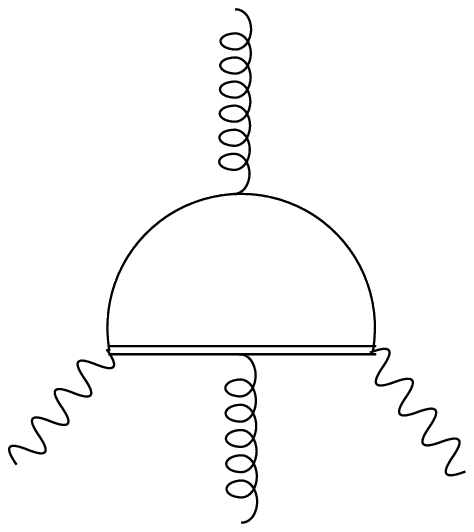}\qquad
\epsfxsize=0.25\textwidth\epsffile{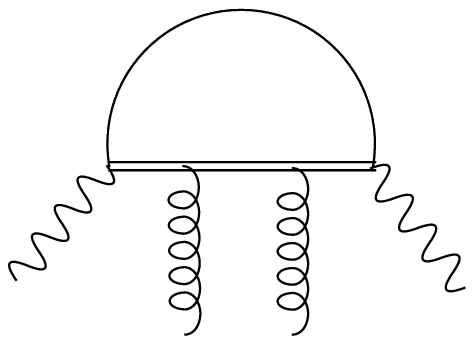}
$$
\vspace*{-20pt}
\caption[]{\small Feynman diagrams of the gluonic contributions. The double
  line denotes the $b$ quark, the photon-like lines the currents in
  the correlation function $\Pi^P_\mu$. The first diagram is
  divergent, the other two are convergent.}\label{fig3}
\end{figure}

In order to obtain the singlet contribution to $\Pi_+^P$, one needs to
calculate the diagrams shown in Fig.~\ref{fig3}.
The projection of the gluon fields onto the DA $\psi_{2;P}^g$ can be
read off Eq.~(\ref{15}). The explicit formula is given in the appendix.
We check the result by
verifying the cancellation of the $\mu_{\rm IR}$-dependent terms as described
above. The relevant term in the quark Gegenbauer moment $a_2$ is
\begin{equation}\label{33}
\Pi^q_+ \sim 18 f_q F(p_B^2,q^2) a_2 \left( 1 + \frac{\alpha_s}{4\pi}
\,\frac{50}{9}\,\ln\,\frac{\mu_{\rm IR}^2}{m_b^2}\right),
\end{equation}
where $F(p_B^2,q^2)$ is a function of $p_B^2$ and $q^2$. The
logarithmic terms in the convolution of the gluonic diagrams of
Fig.~\ref{fig3} with $\psi_{2;P}^g$ read
\begin{equation}
\Pi^P_+ \sim -\frac{10}{9\sqrt{3}}\,\frac{\alpha_s}{4\pi}\, B_2^g f^1_P
\ln\,\frac{\mu_{\rm IR}^2}{m_b^2}\, F(p_B^2,q^2) \,.
\end{equation}
One can easily convince oneself by expressing $f_q$ via
Eq.~(\ref{11}) in terms of
$f^1_\eta$ and $f^1_{\eta'}$, respectively, and inserting (\ref{33})
into (\ref{32}), that the renormalisation-group equation (\ref{20}) is
fulfilled. 

The final LCSR for $f_+^P$ then reads
\begin{equation}\label{35}
e^{-m_B^2/M^2}\,m_B^2 f_B\, \frac{f_+^P(q^2)}{\sqrt{2}} = \int_{m_b^2}^{s_0}
ds\,e^{-s/M^2}\, \frac{1}{\pi}\,{\rm Im}\,\Pi^P_+(s,q^2)\,,
\end{equation}
with the sum-rule specific parameters $M^2$, the Borel parameter, and
$s_0$, the continuum threshold.

\section{Results and Discussion}\label{sec4}
\begin{figure}
$$\epsfxsize=0.44\textwidth\epsffile{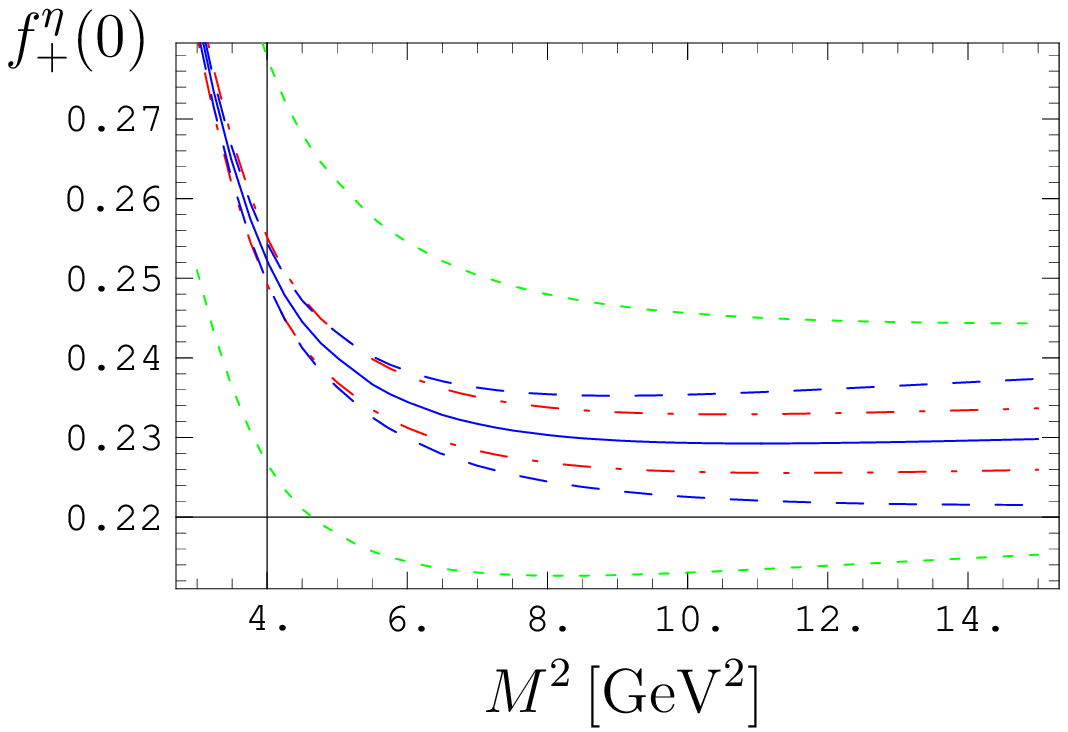}\quad
\epsfxsize=0.44\textwidth\epsffile{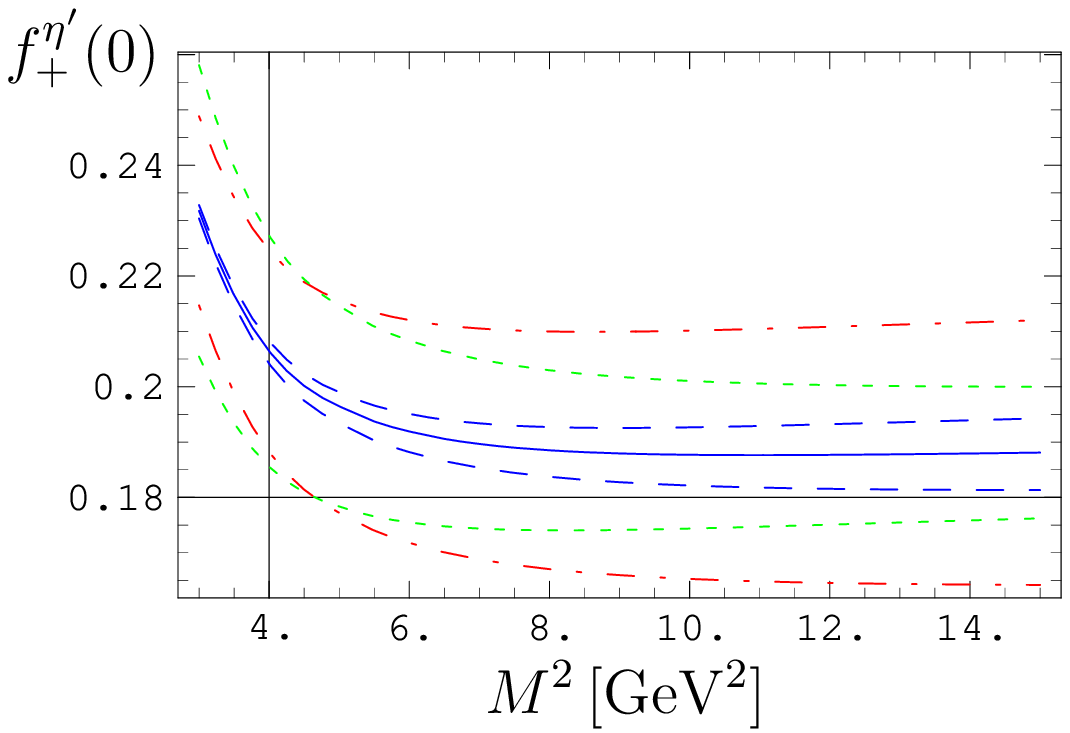}$$
\vskip-17pt
\caption[]{\small
  [Colour online] $f_+^\eta(0)$ (left) and $f_+^{\eta'}(0)$ (right) as a
  function of the Borel parameter $M^2$ and various choices of input
  parameters. Solid curves: central values of input parameters and
  $s_0=34.2\,{\rm GeV}^2$. Long-dashed (blue) curves: $s_0$ varied by
  $\pm 0.7\,{\rm GeV}^2$. Short-dashed (green) curves: $a_2(1\,{\rm
  GeV})$ varied by $\pm 0.15$. Dash-dotted (red) curves: $B_2^g$ varied
  by $\pm 10$.}\label{fig4}
$$\epsfxsize=0.44\textwidth\epsffile{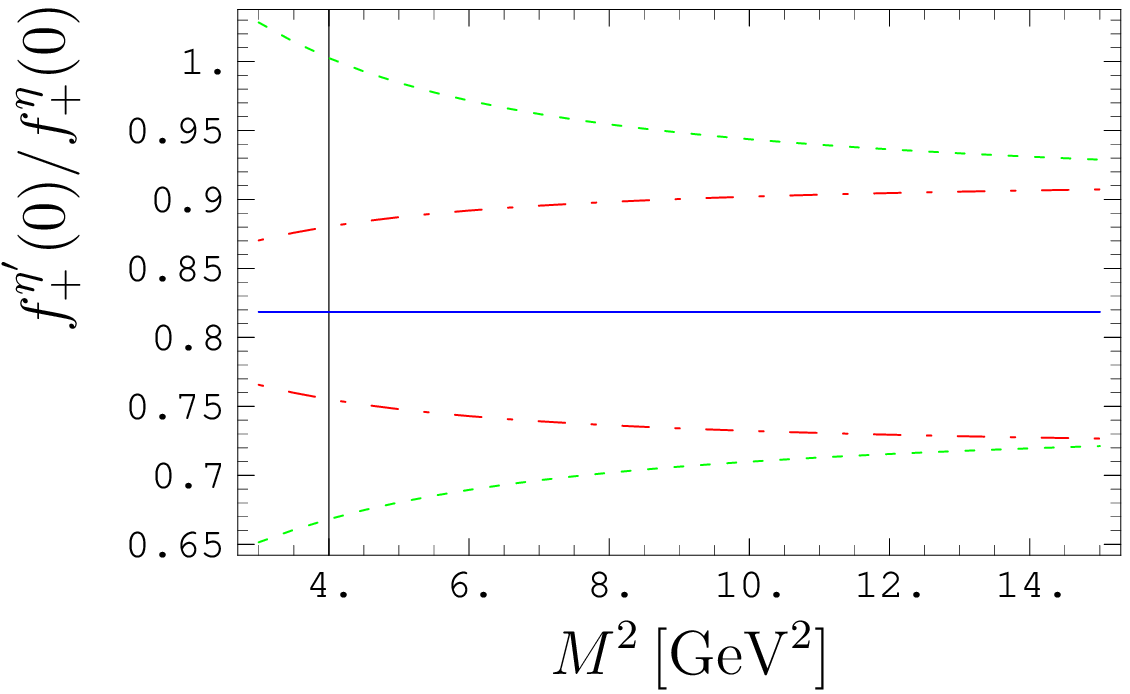}$$
\vskip-17pt
\caption[]{\small
  [Colour online] $f_+^{\eta'}(0)/f_+^\eta(0)$ as a
  function of the Borel parameter $M^2$ and various choices of input
  parameters. Solid (blue) line: central values of input parameters,
  which corresponds to $f_+^{\eta'}(0)/f_+^\eta(0)\equiv \tan\phi =
  0.814$. Dash-dotted (red) curves: $B_2^g$ varied
  by $\pm 10$. Short-dashed (green) curves: $a_2^{\eta,\eta'}(1\,{\rm GeV})$
  varied independently: $a_2^\eta=0.1$, $a_2^{\eta'}=0.4$ and
  $a_2^{\eta'}=0.4$, $a_2^\eta=0.1$.}\label{fig5}
\vskip-8pt
$$\epsfxsize=0.44\textwidth\epsffile{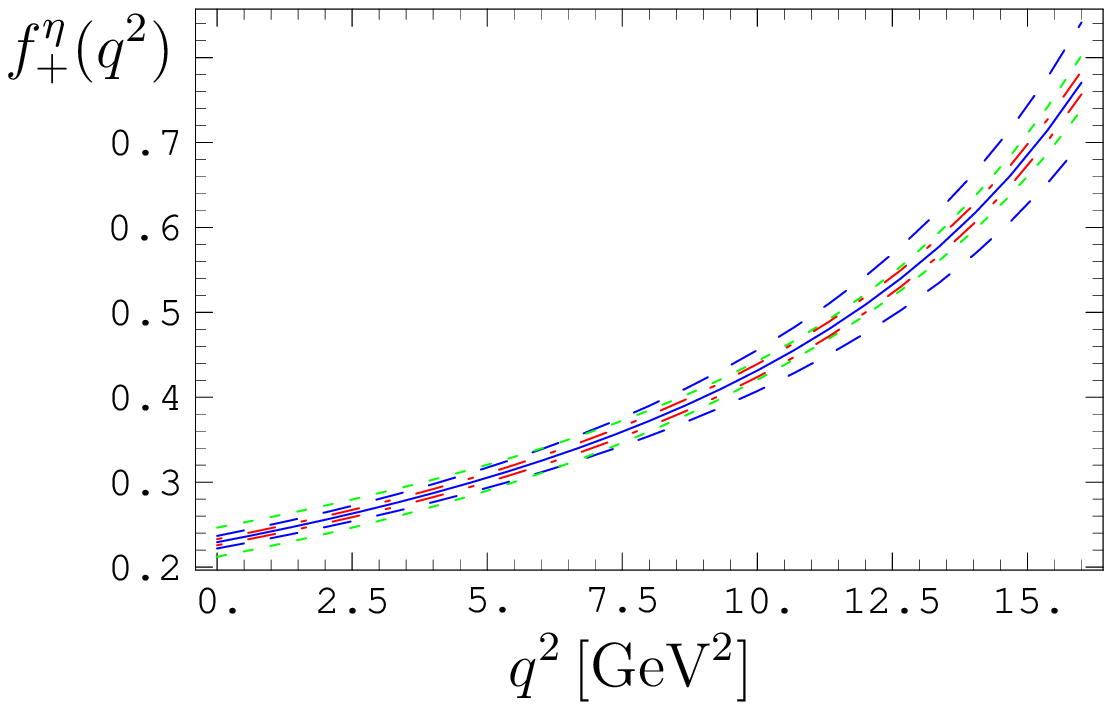}\quad
\epsfxsize=0.44\textwidth\epsffile{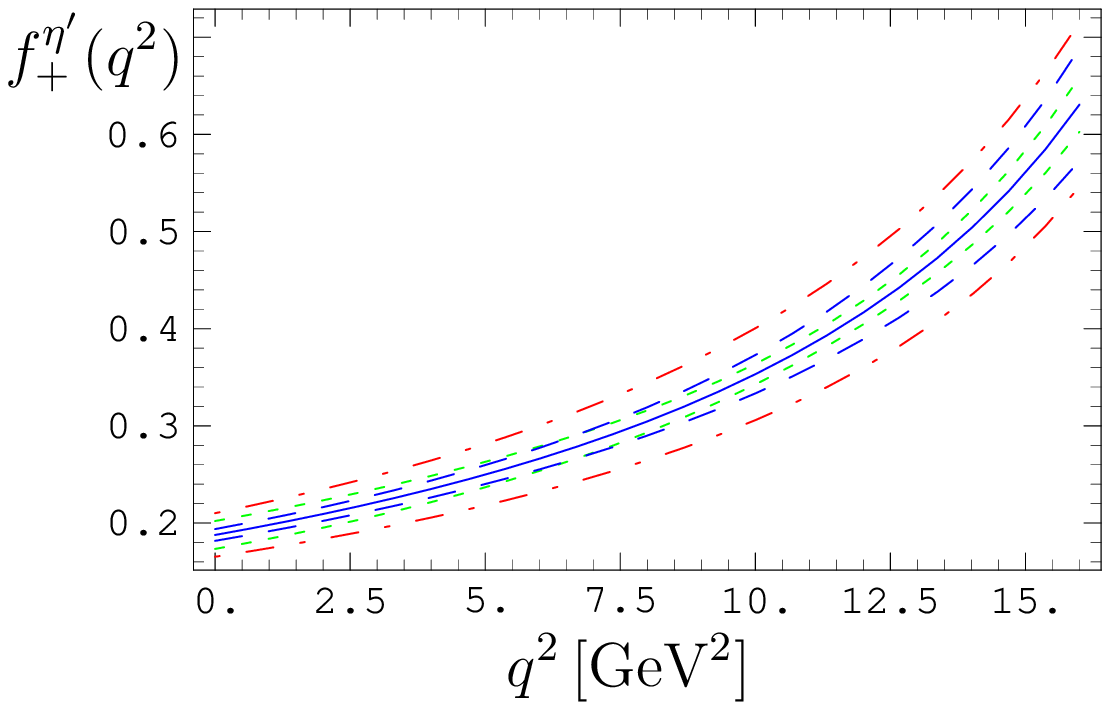}$$
\vskip-17pt
\caption[]{\small [Colour online] $f_+^\eta(q^2)$ (left) and
 $f_+^{\eta'}(q^2)$ (right) as a
  function of the momentum transfer $q^2$ and various choices of input
  parameters. Solid curves: central values of input parameters and 
  $M^2 = 10\,{\rm GeV}^2$, $s_0 = 34.2\,{\rm GeV}^2$. 
  Long-dashed (blue) curves: $s_0$ varied by
  $\pm 0.7\,{\rm GeV}^2$ and $M^2$ by $\pm 4\,{\rm GeV}^2$. 
  Short-dashed (green) curves: $a_2(1\,{\rm
  GeV})$ varied by $\pm 0.15$, $f_q/f_\pi$ by $\pm 0.02$ and $\phi$ by
 $\pm 1^\circ$. Dash-dotted (red) curves: $B_2^g$ varied
  by $\pm 10$.
}\label{fig6}
\end{figure}

Let us now give the results for the form factors. As usual, we replace
$f_B$ in the sum rule (\ref{35}) by its QCD sum rule to $O(\alpha_s)$
accuracy; this reduces the dependence of the results on  $m_b=(4.80\pm
0.05)\,$GeV. 
In Fig.~\ref{fig4}
we plot $f_+^\eta(0)$ and $f_+^{\eta'}(0)$, respectively, as functions
of the Borel parameter $M^2$. The continuum threshold is chosen as
$s_0 = 34.2\,{\rm GeV}^2$, which corresponds to the optimum $s_0$ for
the sum rule for $f_B$ \cite{BZ04}. The factorisation scale $\mu_{\rm
  IR}$ is chosen as intermediate between $m_b$ and an intrinsic
hadronic scale 1~GeV; following our earlier papers, we choose
$\mu_{\rm IR}^2 = m_B^2-m_b^2$. The dependence of
$f_+^{\eta,\eta'}$ on $M^2$ is small in the Borel-window $M^2> 6
\,{\rm GeV}^2$. We estimate the uncertainty in $M^2$ as the variation
of the form factor in the interval $M^2\in [6,14]\,{\rm GeV}^2$. In
Fig.~\ref{fig4}, we also show the dependence of the form factors on
$s_0$ by varying it by $\pm 0.7\,{\rm GeV}^2$; also this dependence is
rather small. The central values of the most relevant 
hadronic input parameters are
$m_b=4.8\,{\rm GeV}$, $a_2^{\eta,\eta'}(1\,{\rm GeV})=0.25$ and
$B^g_2=0$.  As expected, $f_+^\eta(0)$ is not very sensitive to the
singlet contribution parameter $B^g_2$ (red/dashed-dotted curves), but
rather sensitive to the Gegenbauer moment $a_2$ (green/short-dashed
curves). For  $f_+^{\eta'}(0)$, on the other hand, the dependence on
$B^g_2$ is more pronounced than that of $a_2$. Varying all relevant
parameters within their respective ranges, i.e.\ $\Delta m_b = \pm
0.05\,{\rm GeV}$, $\Delta a_2(1\,{\rm GeV})=\pm0.15$ and $\Delta
B^g_2=\pm 20$, as well as all twist-3 and twist-4
parameters within the ranges given in Ref.~\cite{BBL06}, we find
\begin{eqnarray}
f_+^\eta(0) & = & 0.229\pm 0.005(M^2)\pm 0.006(s_0)\pm
0.016(a_2^\eta)\pm 0.007(B^g_2)\pm 0.005(f_q,\phi)\nonumber\\
&&\pm 0.011 ({\rm T3})\pm 0.001({\rm T4}) \pm 0.007(f_B,m_b)\nonumber\\
& = &
0.229\pm 0.024({\rm param.})\pm 0.011({\rm syst.})\,,\\
f_+^{\eta'}(0) & = & 0.188\pm 0.004(M^2)\pm 0.005(s_0)\pm
0.013(a_2^{\eta'})\pm 0.043(B^g_2)\pm 0.005(f_q,\phi)\nonumber\\
&&\pm 0.009({\rm T3})\pm 0.005 ({\rm T4}) \pm 0.006(f_B,m_b)\nonumber\\
& =&
0.188\pm 0.002 B^g_2 \pm 0.019({\rm param.})\pm 0.009({\rm syst.})\,.
\label{fp0}
\end{eqnarray}
The entry labelled T4 also contains an estimate of the possible impact
of the local twist-4 two-gluon matrix elements in (\ref{extra}). For
this estimate, we exploit the fact that the asymptotic DA of the
non-local generalisation of (\ref{extra}) is the same as for the
twist-2 two-quark DA: $6 u (1-u)$.\footnote{This follows from the
  general formula for asymptotic DAs, $u^{2j_1-1} (1-u)^{2j_2-1}$,
  with $j=1/2(l+s)$ the lowest conformal spin of the operator, and $l$
its canonical dimension, $s$ the Lorentz-spin projection. For
$G_{\perp\perp}$, one has $l=2$ and $s=0$ \cite{BBL06}.} We then
assume that the corresponding correlation function is the same as that
for the leading conformal wave in the two-quark twist-2 contribution,
i.e.\ the coefficient in the Gegenbauer moment $a_0=1$, and replace
$a_0$ by $\langle 0 | \alpha_s
G\tilde{G}/(4\pi)|\eta_{q,s}\rangle/(f_{q,s}m_b^2)$. The factor
$1/m_b^2$ comes from the fact that this is a twist-4 effect and hence
suppressed by two powers of $m_b$ with respect to the twist-2
contribution. This is only a
rough estimate, of course, as the true spectral density will be
different. The result in
(\ref{fp0}) shows that for small $B^g_2\approx 2$ both twist-2 and -4
two-gluon effects can indeed be of similar size. In this case,
however, the total flavour singlet contribution to $f_+^{\eta'}$ will
also be small, $\sim 0.008$.
In the third lines, we have added all uncertainties from the input
parameters (param.) in quadrature and the sum-rule specific
uncertainties from $M^2$ and $s_0$ (syst.) linearly. For
$f_+^{\eta'}(0)$, we have displayed the dependence on $B^g_2$
separately. 
Our new result for $f_+^\eta(0)$ is, within errors, in agreement with
our previous one, $f_+^\eta(0) = 0.275\pm 0.036$, obtained in
Ref.~\cite{BZ04}. That for $f_+^{\eta'}(0)$ is new.
Our results agree well with those obtained in Ref.~\cite{Li},
from perturbative QCD factorisation,
$f_+^\eta(0) = 0.208$ and $f_+^{\eta'}(0) = 0.171$, including a
rescaling by a factor $\sqrt{2}$ to bring their definition of the form
factors into agreement with ours. We confirm the finding of
Ref.~\cite{Li} that the range of the singlet contribution to the form
factor estimated in Ref.~\cite{BN02} is likely to be too large, unless
$B_2^g$ assumes extreme values $\sim 40$.

In Fig.~\ref{fig5} we plot the ratio $f^{\eta'}_+(0)/f^{\eta}_+(0)$
as a function of the Borel parameter. In the ratio, many uncertainties
cancel, in particular that on $f_B$. As we have chosen $B_2^g=0$ as
central value, $f^{\eta'}_+(0)/f^{\eta'}_+(0)\equiv \tan\phi = 0.814$ exactly,
see Eq.~(\ref{32}). The figure also illustrates the change of the
result upon inclusion of a non-zero $B^g_2$ (red/dashed-dotted
curves). The ratio is actually rather sensitive to that
parameter. While the dependence on $a_2$ largely cancels when $a_2^\eta$ and
$a_2^{\eta'}$ are set equal, there is a considerable residual dependence on
$a_2^\eta - a_2^{\eta'}\neq 0$ (green/short-dashed curves). While
$|a_2^\eta - a_2^{\eta'}|=0.3$ as illustrated by these curves is
rather unlikely, and would signal very large OZI-breaking
contributions (recall that $a_2^\eta\neq a_2^{\eta'}$ or,
equivalently, $a_2^1\neq a_2^8$ signals the presence of
``wrong-flavour'' contributions to the $\eta_{q,s}$ DAs and is set to 0 in
the QF mixing scheme), one should nonetheless keep in mind that moderate
corrections of this type are not excluded and compete with the
OZI-allowed corrections in $B^g_2$.

Let us now turn to the dependence of the form factors on $q^2$. In
Fig.~\ref{fig6} we show this dependence in the range $0<q^2<16\,{\rm
  GeV}^2$ accessible by LCSRs. Again we display in blue (by long-dashed
curves) the dependence of $f^{\etap}_+(q^2)$ on the sum-rule specific
parameters $M^2$ and $s_0$, the green (short-dashed) curves illustrate
the dependence on $a_2$ and other parameters 
and the red (dash-dotted) ones that on $B^g_2$. We give two
different parametrisations of the form factors, in terms of a sum of
two poles, the so-called BZ parametrisation as given in
Ref.~\cite{BZ04}, and in terms of the BGL parametrisation based on 
analyticity of $f_+$ in $q^2$ \cite{disper}. Both parametrisations are
fitted to the LCSR results in the range $0<q^2<16\,{\rm GeV}^2$, and
can then be used to extrapolate these results to $q^2_{\rm max} =
(m_B-m_{\etap})^2$; this is  possible as both parametrisations include
the essential feature of the $B^*(1^-)$ pole at $q^2=m_{B^*}^2$,
$m_{B^*}=5.33\,{\rm GeV}$, which governs the large-$q^2$ behaviour of $b\to u$
vector-current transitions close to $q^2_{\rm max}$.

The BZ parametrisation reads
\begin{equation}\label{BZ}
f_+(q^2) = f_+(0)\left(\frac{1}{1-q^2/m_{B^*}^2} + \frac{r
  q^2/m_{B^*}^2}{\left(1-q^2/m_{B^*}^2\right)\left(1-
\alpha\,q^2/m_{B}^2\right)} \right),
\end{equation}
with the two shape parameters $\alpha$, $r$ and the
normalisation $f_+(0)$. The BGL parametrisation, on the other hand, is
given by
\begin{eqnarray}
f_+(q^2) & = & \frac{1}{P(q^2) \phi(q^2,q_0^2)}\,\sum_{k=0}^\infty
a_k(q_0^2) [z(q^2,q_0^2)]^k\,,\label{disper}\\
\mbox{with}\quad z(q^2,q_0^2) & = & \frac{\{q_+^2 - q^2\}^{1/2}
- \{q_+^2 - q_0^2\}^{1/2}}{ \{q_+^2 - q^2\}^{1/2}
+ \{q_+^2 - q_0^2\}^{1/2}}\,,\nonumber\\
\phi(q^2,q_0^2) & = & \frac{(q_+^2-q^2)(\sqrt{q_+^2-q_-^2} +
  \sqrt{q_+^2-q^2})^{3/2} (\sqrt{q_+^2-q^2} + \sqrt{q_+^2-q_0^2})}{
(\sqrt{q_+^2} + \sqrt{q_+^2-q^2})^5 (q_+^2-q_0^2)^{1/4}}\,,\nonumber\\
\mbox{and}\quad q_{\pm}^2 & = & (m_B\pm m_{\etap})^2\,.
\end{eqnarray}
The ``Blaschke'' factor
$P(q^2) = z(q^2,m_{B^*}^2)$ accounts for the $B^*$ pole. 
$q_0^2$ is a
free parameter that can be chosen to attain the tightest possible
bounds, and it defines $z(q_0^2,q_0^2) = 0$. One has $|z|<1$ for
$q_0^2<(m_B+m_{\etap})^2$. In the following we choose $q_0^2$ such
that $z(0,q_0^2) \equiv - z(q_-^2,q_0^2)$, i.e.\
$q_0^2=14.14\,{\rm GeV}^2$ for $\eta$ and $10.85\,{\rm GeV}^2$ for
$\eta'$. With these values, $|z|$ becomes minimal:
$|z|<0.13$ for $\eta$ and $|z|<0.08$ for $\eta'$.
The series in (\ref{disper}) provides a systematic expansion in the
small parameter $z$, which for practical purposes has to be truncated
at order $k_{\rm max}$. In this paper, we choose $k_{\rm max}=3$.

\begin{figure}[tb]
$$\epsfxsize=0.46\textwidth\epsffile{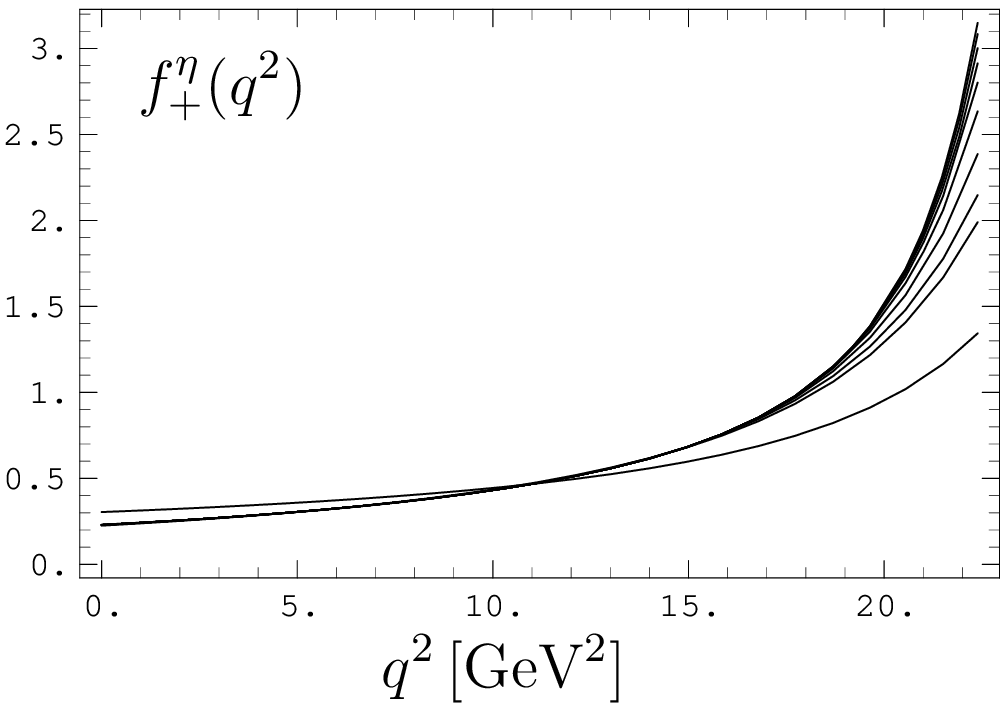}\qquad
\epsfxsize=0.46\textwidth\epsffile{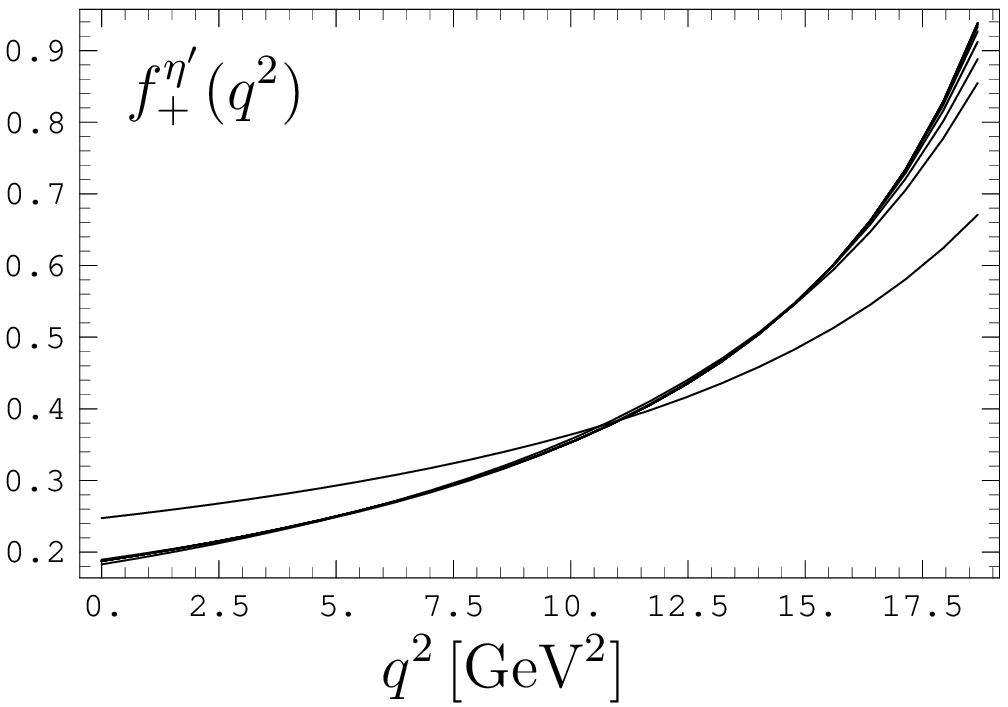}$$
\vskip-16pt
\caption[]{\small $f_+^{\etap}(q^2)$ for central values of input
  parameters, fitted to the BGL parametrisation (\ref{disper}), for
  $0\leq k_{\rm max}\leq 9$.}\label{fig7}
\end{figure}

The advantage of the BZ parametrisation is that it is both 
intuitive and simple: it can be obtained from the dispersion 
relation for $f_+$,
\begin{equation}\label{eq:disper}
f_+^{\etap}(q^2) = \frac{{\rm Res}_{q^2=m_{B^*}^2} f_+(q^2)}{q^2-m_{B^*}^2} +
\frac{1}{\pi} \,\int_{(m_B+m_{\etap})^2}^\infty dt\,\frac{{\rm
    Im}\,f^{\etap}_+(t)}{t-q^2-i \epsilon}\,,
\end{equation}
by replacing the second term on the right-hand side by an effective pole.
However, it cannot easily be extended to include more
parameters. The strength of the BGL parametrisation, on the other
hand, is that the dominant behaviour in $q^2$ close to the pole at
$m_{B^*}^2$ is factored out and the remaining $q^2$-dependence is
organised as a Taylor-series in the small $q^2$-dependent parameter
$z$; the truncation of the series can be adjusted to the accuracy of
the available input parameters. In Fig.~\ref{fig7} we plot
$f_+^{\etap}(q^2)$ parametrised \`{a} la BGL for $0\leq k_{\rm
  max}\leq 9$. Obviously, the parametrisations converge rapidly with
increasing $k_{\rm max}$ and only differ at very
large $q^2$. The impact of this difference on the predicted branching
ratio (\ref{eq:spectrum}) is however only minor, as this region is
phase-space suppressed. In the following, we choose $k_{\rm max}= 3$, 
which ensures that the total predicted branching ratio agrees within
1\% with that obtained for $k_{\rm max}= 9$.

\begin{figure}
$$\epsfxsize=0.5\textwidth\epsffile{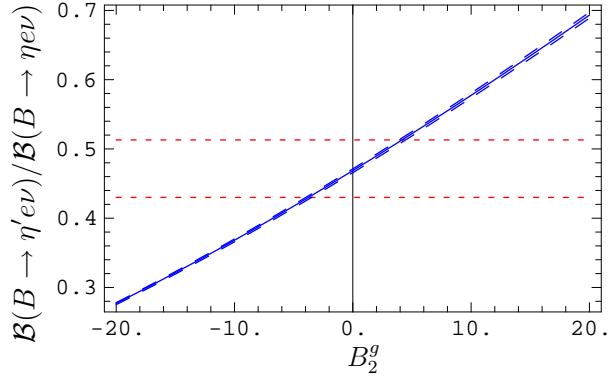}$$
\vskip-16pt
\caption[]{\small [Colour online] The ratio of branching ratios
    $R_{\eta\eta'} = {\cal
    B}(B\to\eta' e\nu)/{\cal B}(B\to\eta e\nu)$ as a function of the
    singlet-parameter $B_2^g$. Solid (blue) curve: central values of
    input parameters and BGL parametrisation with $k_{\rm max}=3$; 
    long-dashed (blue) curves: BGL parametrisations with $k_{\rm max}$ 
    varied by
    $\pm 1$. Short-dashed (red) curves: theoretical uncertainty of
    $R_{\eta\eta'}$ for $B^g_2=0$, for $a_2^{\eta,\eta'}(1\,{\rm GeV})$
    varied independently, as in Fig.~\ref{fig5}.
    }\label{fig8}
\end{figure}

In Tab.~\ref{tab1} we give the best-fit parameters for $f_+^{\etap}$
in the BZ parametrisation, with the small effects of non-zero $B^g_2$
expanded linearly in that parameter. Tab.~\ref{tab2} contains the
corresponding parameters for the BGL parametrisation with $k_{\rm max}
= 3$. 
\begin{table}[tb]
\renewcommand{\arraystretch}{1.3}
\addtolength{\arraycolsep}{3pt}
$$
\begin{array}{c|lll}
 & f_+(0) & \alpha & r\\\hline
\eta & 0.231^{+0.018}_{-0.020} & 0.851^{+0.183}_{-0.492} &
0.411^{-0.030}_{+0.119} \\
\eta' & 0.189^{+0.015}_{-0.016}+ B^g_2\left({}^{+0.002}_{-0.002}\right) 
      & 0.851^{+0.185}_{-0.497}+ B^g_2\left({}^{-0.006}_{+0.008}\right)  
      & 0.411^{-0.031}_{+0.122}+ B^g_2\left({}^{+0.005}_{-0.006}\right)  
\end{array}
$$
\vskip-10pt
\renewcommand{\arraystretch}{1}
\addtolength{\arraycolsep}{-3pt}
\caption[]{\small Parameters for the BZ parametrisation
  (\ref{BZ}). The uncertainty contains all sources of error added in
  quadrature, except for $\eta'$, where the uncertainty in $B^g_2$ is
  approximated by a linear term. The upper (lower) terms represent the
  maximum (minimum) value of the form factor.}\label{tab1}
\renewcommand{\arraystretch}{1.3}
\addtolength{\arraycolsep}{3pt}
$$
\begin{array}{c|ll}
& \eta & \eta'\\\hline
a_0 & \phantom{-}0.0031\pm 0.0003 
    & \phantom{-}0.0018\pm 0.0002\pm 0.00002 B^g_2\\
a_1 & -0.0090\mp 0.0034 & -0.0058\mp0.0016\mp 0.0001 B^g_2\\
a_2 & \phantom{-}0.0243\pm 0.0172 
    & \phantom{-}0.0174\pm0.0166\mp 0.0001 B^g_2\\
a_3 & -0.0908\mp 0.0039 & -0.1189\mp0.0218\pm 0.0016 B^g_2
\end{array}
$$
\vskip-10pt
\renewcommand{\arraystretch}{1}
\addtolength{\arraycolsep}{-3pt}
\caption[]{\small Like Tab.~\ref{tab1}, but for the BGL
  parametrisation (\ref{disper}) with $k_{\rm max}=3$.}\label{tab2}
\end{table}
Finally, in Fig.~\ref{fig8} we show the dependence of the ratio of
branching ratios $R_{\eta\eta'} = {\cal
    B}(B\to\eta' e\nu)/{\cal B}(B\to\eta e\nu)$ on
$B_2^g$. The advantage of this observable is that all hadronic effects
are encoded in the form factors and that $|V_{ub}|$ cancels.
The blue (solid) curve corresponds to the branching ratios obtained
from the central values of input parameters; the dependence of these
predictions on the cut-off in $k$ is very small: the long-dashed
(blue) curves illustrate the dependence on $k_{\rm max}=3\pm 1$. On
the other hand, $R_{\eta\eta'}$ also depends on $a_2^\eta\neq
a_2^{\eta'}$. This dependence is shown by the red (short-dashed)
curves. The conclusion is that large values of $B^g_2$, $|B^g_2|>5$,
can be distinguished from the OZI-breaking parameter
$|a_2^\eta-a_2^{\eta'}|$, once an accurate experimental value of
$R_{\eta\eta'}$ is available, but that for smallish $B^g_2$ and
unknown $|a_2^\eta-a_2^{\eta'}|$ only mutual constraints on these
parameters can be extracted from the data. In this case, as mentioned
before, also twist-4 gluonic DAs can become important. 

To summarise, we have calculated the form factors of $B\to \etap$
semileptonic transitions from QCD sum rules on the light cone,
including the gluonic singlet contributions. We have found that, as
expected, these contributions are more relevant for $f_+^{\eta'}$ than
for $f_+^\eta$ and can amount up to 20\% in the former, 
depending on the only poorly
constrained leading Gegenbauer moment $B^g_2$ of the gluonic twist-2
distribution amplitude of $\etap$. We also found that the form factors
are sensitive to the values of the twist-2 two-quark Gegenbauer
moments $a_2^{\eta,\eta'}$ which, given the uncertainty of independent
determinations, we have set equal to $a_2^\pi$. The ratio
of branching ratios ${\cal B}(B\to\eta' e\nu)/{\cal B}(B\to\eta e\nu)$
is sensitive to both $a_2$ and $B^g_2$ and may be used to constrain
these parameters, once it is measured with sufficient accuracy. The
extraction of $|V_{ub}|$ from these semileptonic decays, in particular
$B\to\eta e\nu$, with negligible singlet contribution, although
possible in principle, at the moment is obscured by the lack of
knowledge of $a_2$. We
would also like to stress that, in the framework of the quark-flavour
mixing scheme for the $\eta$-$\eta'$ system as used in this paper,
$B\to \etap$ transitions probe only the $\eta_q$ component of these
particles. The $\eta_s$ component could be probed directly for
instance in the $b\to s$ penguin transition $B_s\to \etap
\ell^+\ell^-$, although such a measurement would also be sensitive
to new physics in the penguin diagrams.

\section*{Acknowledgments}

G.W.J.\ gratefully acknowledges receipt of a UK PPARC studentship.
This work was supported in part by the EU networks
contract Nos.\ MRTN-CT-2006-035482, {\sc Flavianet}, and
MRTN-CT-2006-035505, {\sc Heptools}.

\appendix
\renewcommand{\theequation}{A.\arabic{equation}}

\section{\boldmath Spectral Density of the two-gluon Contribution to 
$f_+$}\label{app:A}
\setcounter{equation}{0}
The contribution of the twist-2 two-gluon distribution amplitude to
the correlation functions $\Pi_+^\eta$ and $\Pi_+^{\eta'}$,
Eq.~(\ref{32}), is given by
$$\Pi_{+}^{P,1} = \int_{m_b^2}^\infty ds\,\frac{\rho^P_1(s)}{s-p_B^2}$$
with
\begin{eqnarray}
\rho^P_1(s) 
& = & B_2^g a_s f_1^P m_b\, \frac{5}{36\sqrt{3}}\,
\frac{m_b^2-s}{(s-q^2)^5} \, \left\{ 59 m_b^6 + 21 q^6 - 63 q^4 s - 19
q^2 s^2 + 2 s^3\right.
\nonumber\\
&& \hspace*{3cm}\left. + m_b^2 s (164 q^2 + 13 s) - m_b^4 (82 q^2 + 95
s)\right\}
\nonumber\\
&& {} + B_2^g a_s f_1^P m_b\, \frac{5}{6 \sqrt{3}}\, 
\frac{(m_b^2-q^2)(s-m_b^2)}{(s-q^2)^5} \,\{ 5 m_b^4 + q^4 + 3 q^2 s +
s^2 - 5m_b^2 (q^2+s)\}
\nonumber\\
&& \hspace*{3cm}
\times\left\{ 2 \ln\,\frac{s-m_b^2}{m_b^2} - \ln\,\frac{\mu^2}{m_b^2} \right\}.
\end{eqnarray}

\end{document}